\documentclass[journal,transmag]{IEEEtran}
\usepackage{cite}
\usepackage[pdftex]{graphicx}
\usepackage{amsmath}
\usepackage{array}
\usepackage[caption=false,font=footnotesize]{subfig}
\usepackage{dblfloatfix}
\usepackage{mathtools}
\usepackage{physics}
\usepackage{bm}
\usepackage{tikz}
\usepackage{t1enc}
\usepackage{color,soul}
\usepackage{tabu}
\usepackage{lipsum} % for filler text
\usepackage[outdir=./]{epstopdf}

\newcommand{\ba}{\begin{array}}
\newcommand{\ea}{\end{array}}
\newcommand{\be}{\begin{equation}}
\newcommand{\ee}{\end{equation}}
\newcommand{\bd}{\begin{displaymath}}
\newcommand{\ed}{\end{displaymath}}
\newcommand{\bi}{\begin{itemize}}
\newcommand{\ei}{\end{itemize}}
\newcommand{\bn}{\begin{enumerate}}
\newcommand{\en}{\end{enumerate}}
\newcommand{\pa}{\partial}

\newcommand{\mb}{\mathbf}

\begin{document}
\title{Spin-Injection-Generated Shock Waves and Solitons\\ in a
Ferromagnetic Thin Film}


\author{\IEEEauthorblockN{Mingyu Hu\IEEEauthorrefmark{1},
Ezio Iacocca\IEEEauthorrefmark{2}, and 
Mark Hoefer\IEEEauthorrefmark{1}}
\IEEEauthorblockA{\IEEEauthorrefmark{1}Department of Applied Mathematics,
University of Colorado Boulder, Boulder, CO 80309 USA}
\IEEEauthorblockA{\IEEEauthorrefmark{2}Department of Mathematics, Physics, and Electrical Engineering, \\
Northumbria University, Newcastle upon Tyne NE1 8ST, United Kingdom}
% <-this % stops an unwanted space
\thanks{
%Manuscript received December 1, 2012; revised August 26, 2015. 
Corresponding author: M. Hu (email: mingyu.hu@colorado.edu).}}

%% The paper headers
\markboth{CE-07}{}%
%{Shell \MakeLowercase{\textit{et al.}}: Bare Demo of IEEEtran.cls for IEEE Transactions on Magnetics Journals}



% for Transactions on Magnetics papers, we must declare the abstract and
% index terms PRIOR to the title within the \IEEEtitleabstractindextext
% IEEEtran command as these need to go into the title area created by
\IEEEtitleabstractindextext{%
\begin{abstract}
Unsteady nonlinear magnetization dynamics are studied in an easy-plane ferromagnetic channel subject to spin injection at one edge. The model Landau-Lifshitz equation is known to support steady-state solutions, termed dissipative exchange flows (DEFs) or spin superfluids. In this work, by means of numerical simulations and theoretical analysis, we provide a full description of the injection-induced, large-amplitude, nonlinear magnetization dynamics up to the steady state. The dynamics prior to reaching steady state are driven by spin injection, the perpendicular applied magnetic field, the exchange interaction, and local demagnetizing fields. We show that the dynamics take well-defined profiles in the form of rarefaction waves (RW), dispersive shock waves (DSW), and solitons. The combination of these building blocks depends on the interplay between the spin injection strength and the applied magnetic field. A solitonic feature at the injection boundary, signaling the onset of the magnetic "supersonic" condition at the injection edge, rapidly develops and persists in the steady-state configuration of a contact soliton DEF. We also demonstrate the existence of sustained soliton-train dynamics in long time that can only arise in a nonzero applied magnetic field scenario. The dynamical evolution of spin-injection-induced magnetization dynamics presented here may help guide observations in long-distance spin transport experiments.
\end{abstract}

\begin{IEEEkeywords}
ferromagnet, spin injection, nonlinear dynamics, dispersive spin shock wave, soliton
\end{IEEEkeywords}}

\maketitle

\IEEEdisplaynontitleabstractindextext

% \fi
\IEEEpeerreviewmaketitle

\section{Introduction}

%\IEEEPARstart{S}{pin} transport within magnetic materials has been under intense research for potential spintronics applications in information technology. 
%Even though the transport distance has been achieved to be over a micrometer via spin waves \cite{madami2011,cornelissen2015, wesenberg2017,liu2018} by selecting materials with low damping, Yittrium Iron Garnet (YIG) for example, the intensity decays exponentially over the distance \cite{zhang2012,cornelissen2016}. 
\IEEEPARstart{A}{promising} means for long-distance transport of angular momentum is so-called spin superfluidity \cite{sonin2010,takei2014,chen2014,evers2020}. This type of spin transport extends the fluid-like behavior of small-amplitude spin waves, first proposed by Halperin and Hohenberg \cite{halperin1969}, into a large-amplitude regime capable of exhibiting nonlinear waves conveniently analyzed within a dispersive hydrodynamic (DH) framework \cite{iacocca2017}. Here, the magnetization vector $\mathbf{m} = (m_x,m_y,m_z)$ is recast in terms of the longitudinal spin density $n = m_z$ and the magnetic fluid velocity $\mb{u} = -\nabla \arctan (m_y/m_x)$ that is proportional to the spin current, revealing an analogy between magnetodynamics and fluid dynamics. This has been found to be especially beneficial for theoretical studies in the context of spin superfluids and their instabilities \cite{iacocca2020}.
%While some experimental observations have been made on superfluid-like spin transport \cite{stepanov2018,yuan2018}, in order to analytically study large amplitude magnetic textures, the dispersive hydrodynamic (DH) framework has been introduced
%
%The spin transport dynamics can be excited by spin injection of a spin current at material boundaries.
%There are several physical methods proposed to inject spin at boundaries,  theoretically in a ferromagnet \cite{takei2014} through the spin-Hall effect, through spin-transfer torque \cite{iacocca2017symmetry,iacocca2019def,schneider2018,chen2014}, or through the quantum spin-Hall effect experimentally demonstrated in \cite{yuan2018}.
%The steady-state magnetization configurations in a easy-plane ferromagnetic channel subject to spin injection at one end in the absence of external magnetic field have been studied theoretically in \cite{iacocca2019def}. The long-time steady solution, which is the solution to a boundary value problem (BVP), has been identified to be a dissipative exchange flows (DEF) or a contact soliton DEF (CS-DEF), depending on the magnitude of the injected spin current. The emergence of the CS corresponds to the Landau criterion, and in the word of DH, the sonic condition \cite{iacocca2017,iacocca2017symmetry,iacocca2019def}.
The DH representation of the Landau-Lifshitz (LL) equation is an exact transformation and describes the essential physics of a ferromagnet: exchange, anisotropy, and damping, manifested as wave dispersion, nonlinearity, and viscous effects, respectively. In a dispersion-dominated fluid-like medium, large gradients in a physical quantity (e.g. fluid density) can give rise to dispersive shock waves (DSWs) \cite{el2016}. DSWs are expanding, highly oscillatory, nonlinear excitations that realize a coherent transition between two states, the superfluidic, dispersive counterpart to viscous shock waves. Ferromagnets are rich in dispersive phenomena so these dispersive nonlinear wave patterns are expected to arise under the appropriate conditions. Indeed, DSWs have been experimentally observed in the envelope of weakly nonlinear spin waves excited in Yttrium Iron Garnet (YIG) \cite{janantha2017}. 

In this work, we consider the spin transport dynamics excited by spin injection at a material boundary. This can be realized, for example, by the spin Hall effect, which has been experimentally used to detect, e.g., spin waves at long distances \cite{Lebrun2018} and to observe signatures of spin superfluidity \cite{stepanov2018,yuan2018}. The initial condition is a uniform ferromagnetic state with zero velocity $u=0$ everywhere. Subsequently, spin injection at the left boundary is initiated and gradually increases in magnitude, modeled as a hydrodynamic boundary condition (BC) \cite{iacocca2017symmetry} with $|u|$ rising smoothly. 
%This problem setup is equivalent to a piston shock problem, which is a classical problem in fluid dynamics. Moreover, this problem has been studied in the context of superfluids such as Bose-Einstein Condensates (BECs) \cite{hoefer_piston_2008,kamchatnov_flow_2010} where dispersion is a dominating effect and DSWs have been observed as dynamical solutions. In the rest of the paper, we will equivalently refer to the spin injection strength as the piston velocity.
%When dispersion is prominent compare to dissipation and it serves as the main force to regularize discontinuities in the fluid. The rapid change leads to large gradients that result in a new kind of spin shock wave that is an example of a DSW. In addition, DSWs have been identified as solutions to a pure initial value problem (IVP) (the Riemann problem) in a two-component Bose-Einstein condensate \cite{ivanov2017dsw}, whose governing equations are equivalent to the DH form of the no-damping Landau-Lifshitz equation.
In this paper, we present the temporal evolution of magnetization dynamics, described in DH variables, in an easy-plane anisotropic ferromagnetic channel subject to spin injection at one end. Our model incorporates the easy-plane shape anisotropy induced by a thin-film ferromagnetic sample, for which variations in the directions transverse to wave propagation are assumed negligible. This somewhat idealized model has been shown to be quantitatively accurate in realistic micromagnetic simulations in the absence of an externally applied field \cite{iacocca2019def} where the steady-state solutions, dissipative exchange flow (DEF) and contact soliton DEF (CS-DEF), were robustly observed in the presence of nonlocal dipole fields and transverse variations. A DEF, sustained by spin injection, is a stable noncollinear magnetization state that demonstrates spatially diffusing transport of angular momentum, that can be interpreted as a spin current. 
The emergence of a CS corresponds to the magnetic sonic condition \cite{iacocca2017,iacocca2019def}, a valuable concept introduced by the analogy between the DH framework and fluid dynamics. In this work, we identify the stages of development--from spin-injection ramp-up to the steady state--of the magnetization states in the absence and presence of an externally applied magnetic field. The corresponding solution structures within each stage are found to be rarefaction waves (RWs), DSWs, and solitons, depending on the spin injection strength and the applied magnetic field magnitude. In particular, DSWs can only arise when the applied field is nonzero. 
%We observe combinations of these structures in numerical simulations. 
%In the absence of an applied field, the long-time magnetization configuration is a dissipative exchange flow (DEF) or a contact soliton DEF (CS-DEF)  steady state\cite{iacocca2019def}. 
Additionally, a nonzero applied field can give rise to a persistent, propagating, self-interacting soliton-train dynamical solution in long time.

%When a nonzero applied magnetic field is present, the sonic condition depends on the magnitude of the applied field as well because the ferromagnetic nanowire is initially magnetized. We present numerical simulation results of this initial-boundary value problem (IVBP) on the 1 ns timescale (estimated from permalloy). An overview of simulation results is given in Figure \ref{fig:simu_pts}. The solution building blocks include rarefaction waves (RWs), DSWs, and solitons. Detailed discussions are given in later sections.

%\begin{figure*}[t]
%\centering
%\includegraphics[width=7in]{plots/simu_pts_2}
%\caption{Selected simulation locations and identified solution types of the spin injection IVBP transient solutions. Green shade: soliton; Yellow shade: RW; Blue shade: DSW. (|) -- represents a intermediate constant state; (-) -- indicates a composite wave.}
%\label{fig:simu_pts}
%\end{figure*}

\section{Model}
In this work, the magnetization dynamics are effectively modeled by one-dimensional (1D) variations in a planar ferromagnetic channel oriented in the $\hat{\mb{x}}$ direction. 
The governing equation is the non-dimensional LL equation in 1D, given by
\be  \label{eq:LL}
\pa_t \mb{m} = - \mb{m} \cross \mb{h}_{\mathrm{eff}} - \alpha \mb{m} \cross (\mb{m} \cross \mb{h}_{\mathrm{eff}}), \quad x \in (0,L), \ t>0,
\ee
where 
\be \label{eq:heff}
\mb{h}_{\mathrm{eff}} = \pa_{xx} \mb{m} - m_z \hat{\mb{z}} + h_0 \hat{\mb{z}}.
\ee
Here, $\mb{m}$ is the magnetization vector normalized by the saturation magnetization $M_s$. The effective field is $\mb{h}_{\mathrm{eff}} = \Delta \mb{m} - m_z \hat{\mb{z}} + h_0 \hat{\mb{z}}$ , also normalized by $M_s$, consisting of exchange, easy-plane anisotropy, and a constant externally applied magnetic field along the perpendicular-to-plane ($z$) axis, respectively. The Gilbert damping parameter is $\alpha>0$. The non-dimensionalization is achieved by scaling time by $|\gamma| \mu_0 M_s$ and scaling space by $\lambda_{\mathrm{ex}}^{-1}$, where $\gamma$ is the gyromagnetic ratio, $\mu_0$ is the
vacuum permeability, and $\lambda_{\mathrm{ex}}$ is the exchange length. All dimensional quantities quoted in this work are for Permalloy (Py) in which $\gamma=28$~GHz/T, $\mu_0 =4\pi\times10^{-7}$~N/A$^2$, $M_s = 790$~kA/m, $\lambda_{\text{ex}} = 5$~nm, and $\alpha = 0.005$. The DH form is obtained by recasting the LL equation in terms of the hydrodynamic variables
\begin{align*}
\text{spin density: } n &= m_z,\\
\text{fluid velocity: } u &= -\pa_x \Phi = -\pa_x \arctan(m_y/m_x).
\end{align*} 

The spin injection at $x=0$ is modeled as a perfect spin source. At $x=L$, we assume a perfect spin sink with no spin pumping, modeled as a free spin BC. Thus, the BCs are given by
\begin{subequations}
\begin{align}
\pa_x n(x=0,t) = 0, \quad & \pa_x n(x=L,t) = 0, \label{eq:bc_n}\\
u(x=0,t) = u_b(t), \quad & u(x=L,t) = 0, \label{eq:bc_u}
\end{align}
\end{subequations}
where $u_b(t)$ is the time-dependent spin injection strength whose magnitude increases from 0 at $t=0$ to the final intensity $|u_0|$ monotonically and smoothly. 
%This process can be equivalently viewed as accelerating an initially zero-velocity piston. 
In the simulations, we adopt a hyperbolic tangent profile $u_b(t) = \frac{u_0}{2} \left[ \tanh \left( \frac{t - t_0/2}{t_0/10} \right) + 1 \right]$ to model a smooth change in the fluid velocity and thus the rise time $t_0$ is defined as the time where the injection magnitude reaches 99.99\% of its extremum $|u_0|$. In addition, we consider only modulationally stable dynamics \cite{law2001,iacocca2017} by restricting the injection to $|u_0|<1$, so there are no long-wave instabilities.
%However, the value of $u$ can grow beyond $1$ in space as time evolves.
%Initially, the ferromagnetic thin film has $u = 0$ everywhere and is uniformly magnetized by the applied field for $h_0 \leq 1$. 
%It can be viewed as a strip of initially static fluid with a constant density. 
The initial condition (IC) in the DH variables is given by
\begin{subequations}
\begin{align} 
n(x,t=0) &= h_0, \label{eq:ic_n}\\
u(x,t=0) &= 0, \label{eq:ic_u}
\end{align}
\end{subequations}
with $|h_0| < 1$.

The long-wave phase velocities can be derived from the spin-wave dispersion of waves on a uniform hydrodynamic state (UHS), described by spatially uniform spin density and fluid velocity $\bar{n}$ and $\bar{u}$ \cite{iacocca2017}, and are given by
\be 
s_{\pm} = 2 \bar{n} \bar{u} \pm \sqrt{(1-\bar{n}^2)(1-\bar{u}^2)}.
\ee
The current system is identified to be subsonic when $s_- < 0 < s_+$ and supersonic when $s_+ < 0$. In a supersonic system, $|u|$ is larger than the magnetic sound speed $|u_{\text{sonic}}| = \sqrt{(1-\bar{n}^2)/(1+3 \bar{n}^2)}$. In addition, we use the long-wave velocities to predict the dynamical structures that arise for given spin injection and applied field. The temporal evolution of spin-injection-induced dynamics involves three stages,
\bn
\item[1:] Injection rise. If the injection is supersonic, that is when $s_+|_{x=0}<0$, a CS at the injection end is developed within the rise time. The emergence of the CS can be understood as an accumulation of spin current at the injection boundary because the long waves propagate to the left and encounter the boundary. We show in the simulation section that the CS developed at this stage typically persists throughout the dynamical evolution and the steady state. The remaining dynamics can be modeled as effectively damping-free. There are two possible solution types:
\bn
\item[(a)] If $s_+ |_{x=0} < s_+ |_{x=L}$, \textit{expansion} dynamics occur, for example a RW. The solution of this type can be further approximated by the long-wavelength limit.
\item[(b)] If $s_+ |_{x=0} > s_+ |_{x=L}$, \textit{compression} dynamics occur and self-steepening in physical variables takes place. Within the injection time, the self-steepening results in wave-breaking that leads to the formation of a dispersive shock.
\en
One mixed case is also possible with $s_+ |_{x=0} > s_+ |_{x=L}$ at an early time during injection rise and $s_+ |_{x=0} < s_+ |_{x=L}$ later.
\item[2:] Pre-relaxation when the spin injection is maintained at its maximum strength. The dynamics in this stage can be approximated by the conservative limit on times $t_0 < t \sim \frac{1}{\alpha}$ as we will show in simulation. Besides the CS, the rest of the solution structure continues to develop. This stage marks the temporal range where the dispersive hydrodynamics become fully developed, dissipation has not diminished prominent features in the solution structures, and the right boundary has not interacted with the developed dynamics.
\item[3:] Relaxation. In this stage, damping is essential and drives the system to a long-time configuration.
\en
\begin{figure*}[h!]
\centering
\includegraphics[width=5.5in]{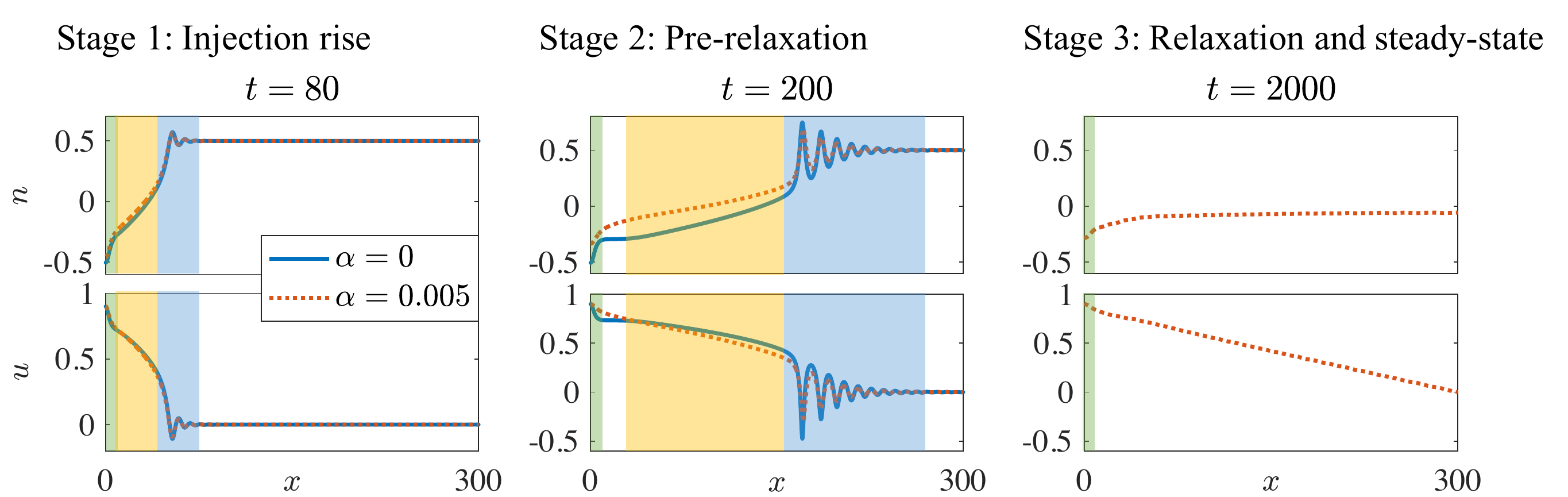}
\caption{An example time evolution of 1D magnetization dynamics excited by spin injection in a planar ferromagnetic channel of length $L = 300$ (1.5~$\mu$m): injection rise time $t_0 = 80$ (2.8 ns), injection strength $u_0 = 0.9$, applied magnetic field $h_0 = 0.5$. Blue solid line: damping coefficient $\alpha = 0$; orange dotted line: $\alpha = 0.005$. Green shade: CS; yellow shade: RW; blue shade: DSW.}
\label{fig:time_evolution}
\end{figure*}
An example of the entire time evolution exhibiting all dynamical structures in terms of the hydrodynamic variables $n$ and $u$ is shown in Fig.~\ref{fig:time_evolution}. The detailed numerical scheme used is presented in the next section. In the example, the ferromagnet length is $L=300$ (1.65~$\mu$m for Py), the injection rise time is $t_0 = 80$ (2.8~ns for Py), the maximum injection strength is $u_0 = 0.9$, and the applied field is $h_0 = 0.5$. 
%Within Stage 1, the injection is found to first satisfy 1(b) (compression) and then (1a) (expansion). During the injection rise, wave-breaking takes place and the onset of a DSW structure appears. In addition, during the injection rise, $s_+|_{x=0}$ becomes negative and thus a CS, in the green (left) shade, is developed at the end of Stage 1. 
During the injection rise in Stage 1, the injection induces both compression and expansion, giving rise to a variety of structures. First, the injection satisfies case 1(b) (compression) which leads to self-steepening (shaded in blue). Then, the injection satisfies case 1(a) (expansion) so that the structure is slowly varying (shaded in yellow). Finally, $s_+|_{x=0}$ becomes negative and a CS develops at the injection boundary (shaded in green).
Within the injection rise, the damped and the undamped solution are almost identical, indicating that the dynamics are dominated by dispersion. 
%In Stage 2, both a RW and a DSW develop and a intermediate constant state develops between the CS and the RW in the undamped solution. The RW, in the yellow (middle) shade, is an expanding long-wave structure. A DSW, in the blue (right) shade, is an expanding, highly-oscillatory, periodic structure with large-amplitude at one edge and diminishing wave amplitude at the other edge \cite{el2016}. 
In Stage 2 at $t=200$ ($\approx7$~ns for Py), the compression and expansion dynamics develop further into well-defined structures. Reading from right to left in the middle panel of Fig.~\ref{fig:time_evolution}: Compression leads to a DSW (shaded in blue), an expanding, highly-oscillatory, and rank-ordered structure with large amplitude at one edge and diminishing wave amplitude at the other edge \cite{el2016}; Expansion gives rise to a RW (shaded in yellow), which is an expanding, non-oscillatory, slowly-varying wave. The CS (shaded in green) remains pinned to the injection site and an intermediate constant state develops between it and the RW.
%\mingyu{We point out that a DSW here does not represent a long-wave supersonic condition as dispersion, as oppose to viscous effects, dominates the dynamics in the time scales of Stage 2.} 
It is worth pointing out that DSWs and RWs have been identified in a two-component Bose-Einstein condensate whose approximate governing dynamical equations are the same as the DH formulation of the dissipationless LL \cite{congy_dispersive_2016,ivanov2017dsw}.
%For the chosen length of the ferromagnetic nanowire, the dynamics at $t=200$ (7 ns for Py) demonstrate the fully developed nonlinear wave structures. 
Here, both DSW and RW persist in the presence of damping (dashed red curves) while the intermediate state between the RW and the CS is lost. Finally, in Stage 3 the damping-driven relaxation process dissipates all oscillations and relaxes the system to the steady-state solution, a CS-DEF. 
%The CS formed within Stage 1 still exists even though its amplitude has been reduced by damping. 
%In the following section, we show simulation results of the fully developed nonlinear dynamics in Stage 2 and the steady-state solution at the end of Stage 3. 

\section{Numerical Simulations}
In this section, we present the simulation results of the magnetization dynamics induced by spin injection in a planar  ferromagnetic channel. We solve the LL equation~\eqref{eq:LL} subject to the BCs
\be
\pa_x \mb{m}(x=0,t) = u_b(t) \mb{m}(x=0,t) \cross \hat{\mb{z}}, \ \pa_x \mb{m}(x=0,t) = 0, \label{eq:bc_m}
\ee
where the spin injection BC is of a Robin (mixed) type and the free spin BC is of the Neumann type \cite{takei2014}. The IC is given by
\begin{subequations}
\begin{align} 
m_x(x,t=0) &= \sqrt{1-h_0^2}, \label{eq:ic_mx}\\
m_y(x,t=0) &= 0, \label{eq:ic_my}\\
m_z(x,t=0) &= h_0. \label{eq:ic_mz}
\end{align}
\end{subequations}
This initial-boundary value problem (IBVP) is solved using the method of lines: the right-hand-side of Eq.~\eqref{eq:LL} is spatially approximated by the sixth-order centered finite difference method and the result serves as an approximation to the time derivative at the current time; then the IC is time-stepped discretely using the \texttt{MATLAB} built-in initial value problem (IVP) solver \texttt{ode23}. The spin injection and free spin BCs are implemented using 6th-order local extrapolation polynomials with ghost points \cite{gibou2005fourthFD} to maintain the order of accuracy and smoothness in the solution near the boundaries. 
%We consider material parameters for a Permalloy with $M_s = 790$ kA/m, an exchange length $\lambda_{\text{ex}} = 5$ nm, and a damping coefficient $\alpha = 0.005$. 
%We simulate a 1D nanowire of non-dimensional length $L = 300$ (dimensional: $300 \lambda_{\text{ex}}$, approximately 1.65 $\mu$m). The rise time is chosen to be $t_0 = 80$, which is non-dimensionalized by $|\gamma| \mu_0 M_s$ and is approximately 0.4 ns. 
During the rise time, the spin injection changes slowly enough so as to reduce additional oscillations. With this numerical model, we explore the unsteady magnetization dynamics in the absence and presence of an externally applied magnetic field. We show results at Stage 2, where the nonlinear solutions are fully developed, and Stage 3, where dissipation dominates the solution.

\subsection{Zero Applied Field $h_0 = 0$}
\begin{figure}[h!]
\centering
\includegraphics[width=3.5in]{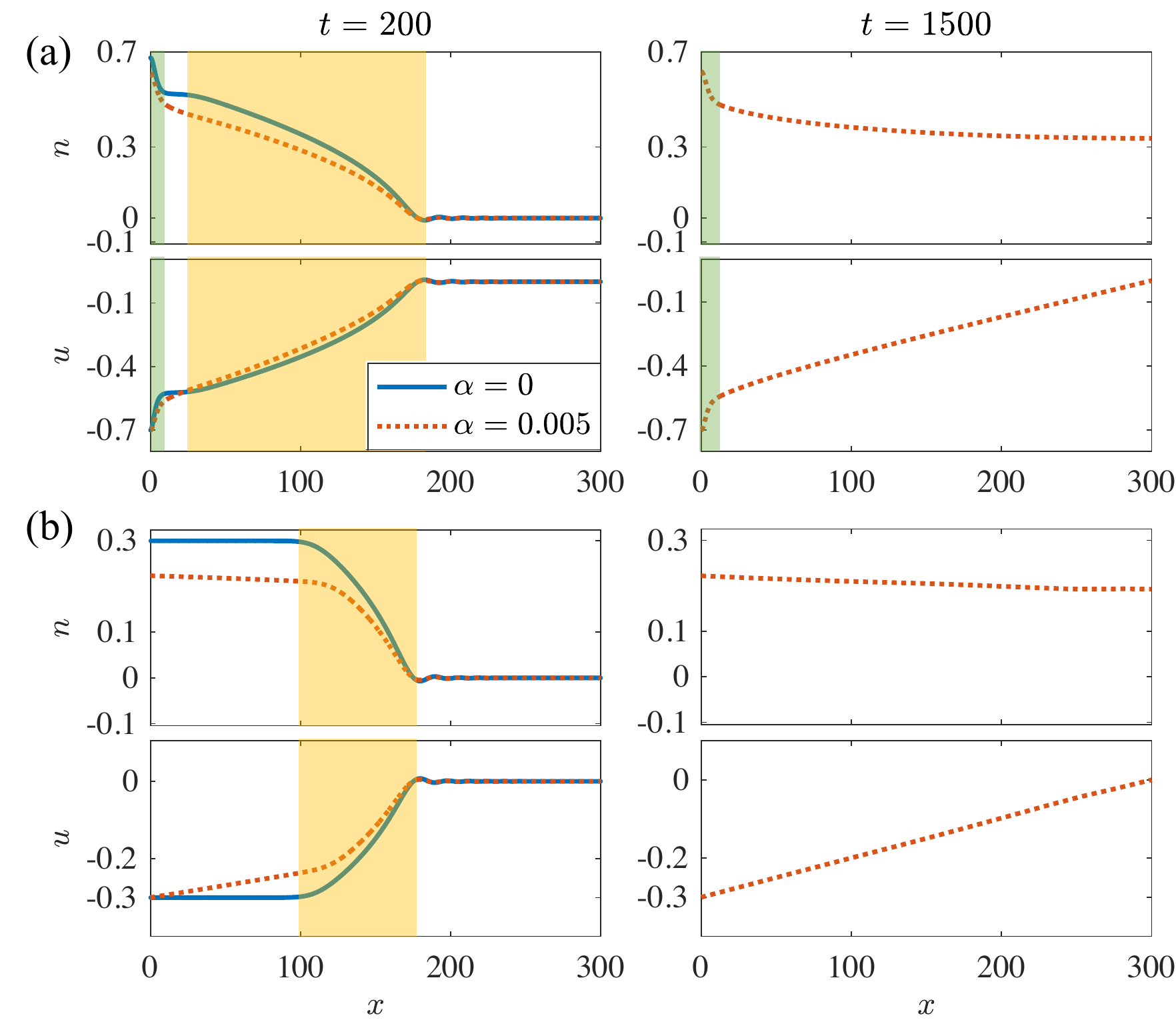}
\caption{Simulation results of Stage 2 (left panels) and Stage 3 (right panels, 52.5 ns) of the time evolution of spin-injection-induced dynamics in a planar ferromagnetic channel of length $L = 300$ (1.5~$\mu$m) with zero applied field $h_0 = 0$. The spin injection intensities are (a) $u_0 = -0.7$, (b) $u_0 = -0.3$. }
\label{fig:damped_undamped_h0_0}
\end{figure}

Fig.~\ref{fig:damped_undamped_h0_0} shows the numerical simulation results for the zero field case. We only discuss the negative injection results here since the solutions for positive injection strength have the opposite signs, due to the symmetry $x \rightarrow -x$ and $u \rightarrow -u$ when $h_0 = 0$. 

We start our discussion with $u_0 = -0.7$ solutions, shown in Fig.~\ref{fig:damped_undamped_h0_0}(a). In Stage 1, it is found that the long-wave velocities satisfy $s_+ |_{x=0} < s_+ |_{x=L}$ (case 1(a)) throughout the injection rise time. Hence, only expansion dynamics arise. In addition, $s_+|_{x=0}<0$ and thus the system becomes supersonic and a CS is formed during this stage. In Stage 2, shown in the left panels of Fig.~\ref{fig:damped_undamped_h0_0}(a), a RW (shaded in yellow) develops and expands while the CS (shaded in green) stays stationary at the injection end. We term the undamped solution a CS|RW, with "|" denoting the intermediate constant state in between. 
%Within Stage 2, our simulations show that the damped configuration still retains the CS and RW features.
%while the intermediate constant state is lost. 
We point out that the small-amplitude oscillations at the leading edge of the RW are caused by the injection rise dynamics and are not a DSW solution \cite{congy_dispersive_2016,el2016}.
At the end of Stage 3, the damped solution reaches a CS-DEF steady-state configuration. 
%The CS, sustained by supersonic injection, survives the relaxation process. 
Notice that the RW has decayed into the DEF.

The $u_0 = -0.3$ solution is shown in Fig.~\ref{fig:damped_undamped_h0_0}(b). Throughout Stage 1 of the injection rise, it is found that $s_+ |_{x=0} < s_+ |_{x=L}$ (case 1(a)). Hence, only expansion dynamics (shaded in yellow) are identified in Stage 2,  shown in the left panels in Fig.~\ref{fig:damped_undamped_h0_0}(b). The steady-state solution after relaxation is identified to be a DEF for this subsonic injection.

%Our simulations for zero applied field demonstrate that the nonlinear  dynamics in Stage 2 of the full time evolution can be approximated by the undamped solutions and they eventually reach the previously-studied steady-state solutions, DEF for subsonic injection and CS-DEF for supersonic injection. The CS is the signature phenomenon when the injection is supersonic: it is developed during Stage 1 of the injection rise time and persists over the damping-driven relaxation process until the steady-state configuration. 

\subsection{Uniform Perpendicular Applied Field $h_0 \neq 0$}
\begin{figure}[h!]
\centering
\includegraphics[width=3.5in]{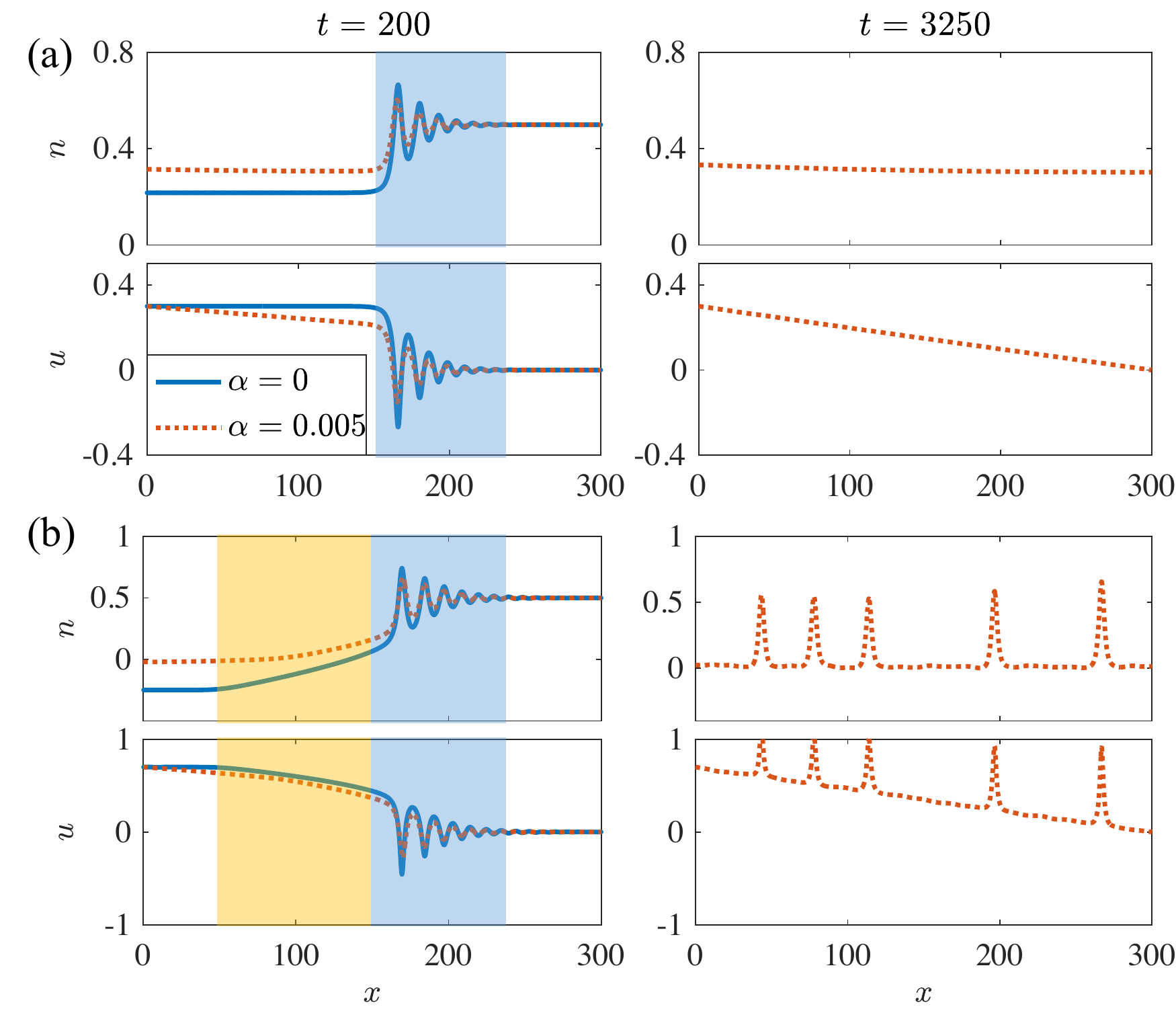}
\caption{Simulation results of Stage 2 (left panels) and Stage 3 (right panels, 114~ns) of the time evolution of spin-injection-induced dynamics in a planar ferromagnetic channel of length $L = 300$ (1.5~$\mu$m) with constant applied field $h_0 = 0.5$ along the perpendicular $z$-axis. The spin injection intensities are (a) $u_0 = 0.3$, (b) $u_0 = 0.7$. An additional supersonic simulation result with $u_0 = 0.9$ is shown in Fig.~\ref{fig:time_evolution}.}
\label{fig:stlt_h0_nz}
\end{figure}

Fig.~\ref{fig:time_evolution} and \ref{fig:stlt_h0_nz} show the numerical simulation results when $h_0 = 0.5$. 
It is found that the dynamical and steady-state solutions with negative injections have the same solution structures as the zero field case discussed earlier: a RW for subsonic injection relaxes to a DEF, and a CS|RW for supersonic injection relaxes to a CS-DEF. Therefore, we only further discuss the positive injection results here. In addition, we only present the solutions for a positive field (in the $+z$ direction) because there exhibits an odd symmetry in the $u_0-h_0$ plane, a generalization of the symmetry about $u_0 = 0$ in the zero field case. We also point out that relaxation (Stage 3) requires longer time to achieve a long-time configuration because of the large amplitude in a DSW.

We start the discussion with the small injection strength $u_0 = 0.3$. During Stage 1, it is found that $s_+ |_{x=0} > s_+ |_{x=L}$ (case 1(b)) throughout. Thus compression dynamics are generated immediately and self-steepening is expected to lead to wave-breaking that results in a highly-oscillatory DSW. The simulation shown in the left panel of Fig.~\ref{fig:stlt_h0_nz}(a) confirms this prediction (shaded in blue). The DSW reveals the dispersion-dominated dynamics that are characteristic of ferromagnets on short enough time scales. In addition, $s_-|_{x=0} < 0 < s_+|_{x=0}$ for $u_0 = 0.3$, and hence the magnetic "flow" is always subsonic without a CS. After relaxation, the steady-state solution is a DEF, shown in the right panel of Fig.~\ref{fig:stlt_h0_nz}(a).

%\begin{figure*}[t!]
%\centering
%\includegraphics[width=5in]{plots/solitons_on_DEF_contours}
%\caption{Long-time dynamics for $u_0 = 0.7$, $h_0 = 0.5$, a traveling soliton train on DEF. (a) Space-time contour of long-time dynamics from $t=3250$ to $t=3500$ before the left-traveling soliton train interact with the injection boundary on the left. (b) Space-time contour of long-time dynamics from $t=3250$ to $t=5000$ revealing a full cycle of the soliton train traveling left, being reflected by the injection boundary,  self-interacting, traveling right, and being reflected by the spin sink boundary.}
%\label{fig:soli_on_def}
%\end{figure*}

\begin{figure}[t!]
\centering
\includegraphics[width=3in]{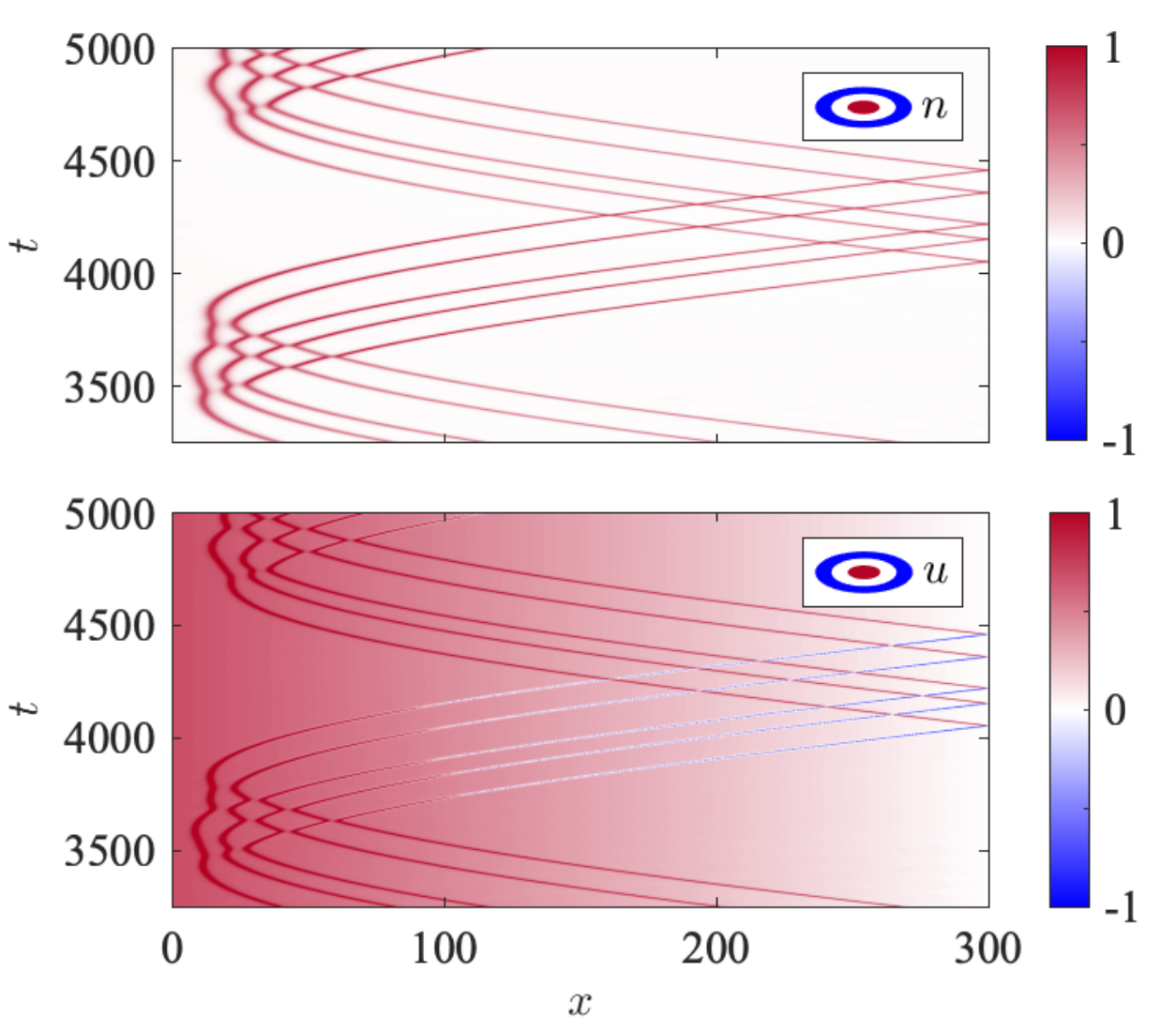}
\caption{Space-time contour of long-time dynamics for $u_0 = 0.7$, $h_0 = 0.5$, a traveling soliton train on DEF. This contour plot demonstrates a full cycle of the soliton train traveling left, being reflected by the injection boundary,  self-interacting, traveling right, and being reflected by the spin sink boundary.}
\label{fig:soli_on_def}
\end{figure}

With moderate injection $u_0 = 0.7$, during Stage 1, it is found that $s_+ |_{x=0} > s_+ |_{x=L}$ (case 1(b)) at first and then $s_+ |_{x=0} < s_+ |_{x=L}$ (case 1(a)) thereafter. Therefore, compression dynamics are induced immediately and a DSW will emerge as a result of self-steepening and wave-breaking. This is followed by expansion dynamics, manifesting in a RW. Both of these structures are fully developed by Stage 2, shown in the left panel of Fig.~\ref{fig:stlt_h0_nz}(b). The DSW (shaded in blue) is observed directly adjacent to the RW (shaded in yellow). The left edge of the DSW travels at the same speed as the right edge of the RW. Hence, this DSW is the analog of a contact discontinuity in classical fluids and is termed a contact DSW (cDSW) \cite{congy_dispersive_2016,ivanov2017dsw}. Thus, we term the pre-relaxation solution a RW-cDSW composite wave. 
Similarly, the supersonic solution with  $u_0 = 0.9$ shown in Fig.~\ref{fig:time_evolution} is a CS|RW-cDSW. 
Throughout Stage 1, $s_-|_{x=0} < 0 < s_+|_{x=0}$ so the solution remains subsonic. In Stage 3, a new long-time configuration is observed. In Fig.~\ref{fig:stlt_h0_nz}(b) right panel and Fig.~\ref{fig:soli_on_def}, the background mean flow can still be identified as a DEF steady-state, but there is additionally a train of solitons on top. These solitons are dynamic, as they travel back and forth and interact with each other within the ferromagnet. A complete cycle of the periodic motion of the soliton train is shown in Fig.~\ref{fig:soli_on_def}. It is also observed that vacuum states, where $|n| = 1$ and $u$ switches sign, can be reached after the solitons are reflected by the spin injection boundary.
The simulation in Fig.~\ref{fig:soli_on_def} indicates that these solitons are amplified to reach vacuum after being reflected by the spin injection left boundary and then decrease in amplitude to fall back from vacuum through the spin sink at the right boundary. 
These distinct long-time dynamics call for a different analytical description than those for the DEF/CS-DEF steady states.

\section{Discussion and Conclusion}
In this paper, we described the time evolution of magnetization dynamics induced by spin injection at one edge of an effective easy-plane ferromagnetic channel. Our analysis utilizes the DH framework, which provides a fluid analogy of ferromagnetism. We use the long-wave velocities during the injection rise stage to predict the solution structures that are verified qualitatively by numerical simulations.
%We identify three stages in the dynamical evolution until a steady state: the spin injection stage, the pre-relaxation stage, and the relaxation stage until the steady state. 
If the injection is supersonic, a CS at the injection site completes development by the end of the rise time and lives through the entire time evolution. This signature feature indicates that there is a saturation limit of angular momentum a thin-film ferromagnet can support through nonlinear textures. During pre-relaxation (Stage 2), highly-oscillatory dispersive wave structures, such as a DSW and a RW-cDSW, arise only when the applied magnetic field is non-zero. A more detailed theoretical description is needed to clarify the interplay between the spin injection strength and the externally applied field magnitude that gives rise to these structures in order to identify their salient features.
%We emphasize that the time frame of the pre-relaxation stage  depends on the nanowire length and the damping coefficient: longer channel and smaller damping coefficient yield a longer duration of the pre-relaxation dynamics. 
After the relaxation process, other than a DEF or a CS-DEF steady-state configuration, our numerical simulation reveals a dynamical long-time solution of a train of traveling, interacting solitons on a DEF profile in a subsonic scenario. The conditions required for this novel long-time behavior and its mechanism are under further investigation. We also point out that the presented simulation results are obtained under ideal conditions: a defect-free ferromagnet with only local dipole fields, a perfect spin source at one boundary, and a perfect spin sink at the other. 
%While only the long-wave stability is ensured in the initial condition, the system may undergo other instabilities and especially in a real experimental setting, any disturbance could lead to corruption of the predicted solution structures.
%The long-time steady-state solutions in the case of nonzero applied field also points to new theoretical approaches beyond the DEF/CS-DEF analytical descriptions. 
Nevertheless, the predicted time evolution of magnetization dynamics suggests new features to look for in an experimental realization of microscopic spin transport in a ferromagnet that can be detected in the nanosecond regime.

% or


%\appendices
%% you can choose not to have a title for an appendix
%% if you want by leaving the argument blank
%\section{}

\section*{Acknowledgment}
The authors acknowledge funding from the National Institute of Standards and Technology Professional Research Experience Program and the U.S. Department of Energy, Office of Science, Office of Basic Energy Sciences under Award No. DE-SC0018237.

\ifCLASSOPTIONcaptionsoff
  \newpage
\fi



\bibliographystyle{IEEEtran}
\bibliography{2021_intermag_transmag_HU_arxiv.bib}

%\vfill


\end{document}